\documentstyle[12pt,epsf]{article}

\begin{document}


\begin{center}

 {\Large \bf
\vskip 7cm
\mbox{Exclusive Double Diffractive Higgs}
\mbox{Boson Production at LHC.}
}
\vskip 3cm

\mbox{Petrov~V.A. and Ryutin~R.A.}

\mbox{{\small Institute for High Energy Physics}}

\mbox{{\small{\it 142 281} Protvino, Russia}}

 \vskip 1.75cm
{\bf
\mbox{Abstract}}
  \vskip 0.3cm

\newlength{\qqq}
\settowidth{\qqq}{In the framework of the operator product  expansion, the quark mass dependence of}
\hfill
\noindent
\begin{minipage}{\qqq}
Exclusive double diffractive (EDD) Higgs boson production is analyzed in the framework of 
the Regge-eikonal approach. Total and differential cross-sections 
for the process $p+p\to p+H+p$ are calculated. Experimental 
possibilities to find Higgs boson at LHC are discussed.

\end{minipage}
\end{center}


\begin{center}
\vskip 0.5cm
{\bf
\mbox{Keywords}}
\vskip 0.3cm

\settowidth{\qqq}{In the framework of the operator product  expansion, the quark mass dependence of}
\hfill
\noindent
\begin{minipage}{\qqq}
Diffractive Higgs Boson Production -- Pomeron -- Regge-Eikonal model
\end{minipage}

\end{center}

\setcounter{page}{1}
\newpage


\section{Introduction}

LHC collaborations aimed at working in low (TOTEM)- and high (CMS)-$p_T$
regimes related to typical undulatory (diffractive) and corpuscular (point-like) 
behaviours of the corresponding cross-sections may offer a very exciting possibility to
observe an interplay of both regimes~\cite{i1}. In theory the "hard part" can be
(hopefully) treated with perturbative methods whilst the "soft" one is definitely
nonperturbative.

 Below we give an example of such an interplay: exclusive Higgs boson production by diffractively
scattered protons, i.e. the process $p+p\to p+H+p$, where $+$ means also a rapidity gap.

 This process is related to the dominant amplitude of exclusive two-gluon 
production. Driving mechanism of the diffractive processes is the Pomeron. Data on the total
cross-sections demands unambiguosly for the Pomeron with larger-than-one intercept, thereof
the need in "unitarisation". 

As will be seen below the detection of the Higgs boson (in the $b\bar{b}$ mode) at LHC in the double
diffractive regime looks fairly possible.

\section{Calculations}

In Figs.~1,2 we illustrate in detail the process
$p+p\to p+H+p$. Off-shell proton-gluon amplitudes in Fig.~2 are treated by the method developed
in Ref.~\cite{i2}, which is based on the extension of Regge-eikonal approach, and 
succesfully used for the description of the HERA data~\cite{i3}.   

 The amplitude of the process $p+p\to p+H+p$ consists of two parts (see~Fig.~2). The "hard" 
part $F$ is the usual gluon-gluon fusion process 
calculated in the Standart Model\cite{SMHiggs}. "Soft" amplitudes $T_{1,2}$ are obtained 
in the Regge-eikonal approach.
  
 We use the following kinematics, which corresponds to the double Regge limit. It is convenient
to use light-cone components $(+,-;\bot)$. The components of momenta of the hadrons in Fig.~2 are
\begin{eqnarray}
\label{momenta}
p_1&=&\left( \sqrt{\frac{s}{2}},\frac{m^2}{\sqrt{2s}}, {\bf 0}\right)\\
p_2&=&\left( \frac{m^2}{\sqrt{2s}},\sqrt{\frac{s}{2}}, {\bf 0}\right)\nonumber\\
p_1^{\prime}&=&\left( (1-\xi_1)\sqrt{\frac{s}{2}},\frac{{\bf\Delta}_1^2+m^2}{(1-\xi_1)\sqrt{2s}}, -{\bf\Delta}_1\right)\nonumber\\
p_2^{\prime}&=&\left( \frac{{\bf\Delta}_2^2+m^2}{(1-\xi_2)\sqrt{2s}},(1-\xi_2)\sqrt{\frac{s}{2}}, -{\bf\Delta}_2\right)\nonumber\\
q&=&\left( q_+, q_-, {\bf q}\right)\;{,}\nonumber\\
q_1&=&q+p_1-p_1^{\prime}=q+\Delta_1\;{,}\nonumber\\
q_2&=&-q+p_2-p_2^{\prime}=-q+\Delta_2\;{,}\nonumber
\end{eqnarray}
$\xi_{1,2}$ are fractions of protons momenta carried by gluons. For two-dimensional 
transverse vectors we use boldface type. From the above notations we can
obtain the relations:
\begin{eqnarray}
\label{notations}
t_{1,2}\;\;=\;\;\Delta_{1,2}^2 &\simeq& -\frac{{\bf\Delta}_{1,2}^2(1+\xi_{1,2})+\xi_{1,2}^2m^2}{1-\xi_{1,2}}\;\;\simeq\;\; -{\bf\Delta}_{1,2}^2\;{,}\;\; \xi_{1,2}\to 0\\
\cos\phi_0&=&\frac{{\bf\Delta}_1{\bf\Delta}_2}{|{\bf\Delta}_1||{\bf\Delta}_2|}\nonumber\\
M_H^2&\simeq& \xi_1\xi_2s+t_1+t_2-2\sqrt{t_1t_2}\cos\phi_0\nonumber\\
(p_1+q)^2&\simeq&m^2+q^2+\sqrt{2s}q_-=s_1\nonumber\\
(p_2-q)^2&\simeq&m^2+q^2-\sqrt{2s}q_+=s_2\; .\nonumber
\end{eqnarray}
Physical region of diffractive events with two rapidity gaps is defined by the following  
kinematical cuts:
\begin{equation}
\label{tlimits}
0.01\; GeV^2\le |t_{1,2}|\le 1\; GeV^2\;{,} 
\end{equation}
\begin{equation}
\label{xlimits}
\xi_{min}\simeq\frac{M_H^2}{s \xi_{max}}\le \xi_{1,2}\le \xi_{max}=0.1\;,
\end{equation} 
\begin{eqnarray}
\label{klimits}
\left(\sqrt{-t_1}-\sqrt{-t_2}\right)^2\le\!\!\!&\kappa&\!\!\!\le\left(\sqrt{-t_1}+\sqrt{-t_2}\right)^2\\
\kappa=\xi_1\xi_2s\!\!\!&-&\!\!\! M_H^2\ll M_H^2\nonumber
\end{eqnarray}
We can write the relations in terms of $y_{1,2}$ and $y_H$ (hadron and Higgs boson rapidities correspondingly):
\begin{eqnarray}
\label{raplimits}
&\xi_{1,2}&\!\!\!\!\simeq\frac{M_H}{\sqrt{s}}e^{\pm y_H}\; ,\\
&|y_H|&\!\!\!\!\le y_0=\ln\left(\frac{\sqrt{s}\xi_{max}}{M_H}\right)\; ,\nonumber\\
y_0\simeq 2.5\!\!\!&\mbox{\rm for}&\!\!\!\sqrt{s}=14\; TeV\;{,}\; M_H=115\; GeV\; , \nonumber\\
&|y_{1,2}|&\!\!\!\!=\frac{1}{2}\ln\frac{(1-\xi_{1,2})^2s}{m^2-t_{1,2}}\ge 9\nonumber
\end{eqnarray}

 The calculation of the amplitude is based on the nonfactorized scheme. The 
contribution of the diagram depicted in Fig.~2
is obtained by integrating over all internal loop momenta. It was 
shown in~\cite{Collins}, that the leading contribution arises from the region of the
integration, where momentum $q$ is "Glauber-like", i.e. of the order 
$(\mbox{k}_+m^2/\sqrt{s},\mbox{k}_-m^2/\sqrt{s},\mbox{\bf k}m)$, where k's are of the 
order $1$. The detailed consideration of the loop integral like
\begin{equation}
\int\frac{d^4 q}{(2\pi)^4}\frac{f(q,p_1,p_2,\Delta_1,\Delta_2)}{(q^2+i 0) (q_1^2+i 0) (q_2^2+i 0)}
\end{equation}
shows that the main contribution comes from the poles 
\begin{eqnarray}
&q_1^2&\!\!\!=\sqrt{2s}\xi_1q_{-}-{\mbox{\bf q}_1}^2 = 0\; ,\nonumber\\ 
&q_2^2&\!\!\!=-\sqrt{2s}\xi_2q_{+}-{\mbox{\bf q}_2}^2 = 0\; .\nonumber
\end{eqnarray}
In this case
\begin{equation}
\label{qvalue}
q=\left(-\frac{{\mbox{\bf q}_2}^2}{\xi_2\sqrt{2s}},\; \frac{{\mbox{\bf q}_1}^2}{\xi_1\sqrt{2s}}, \mbox{\bf q}
\right)\;{,}
\end{equation}
where 
\begin{eqnarray}
&{\mbox{\bf q}_1}^2&\!\!\!\!={\mbox{\bf q}}^2+{{\bf\Delta}_1}^2+2|\mbox{\bf q}||{\bf\Delta}_1|\cos(\phi+\frac{\phi_0}{2})\;{,}\nonumber\\
&{\mbox{\bf q}_2}^2&\!\!\!\!={\mbox{\bf q}}^2+{{\bf\Delta}_2}^2-2|\mbox{\bf q}||{\bf\Delta}_2|\cos(\phi-\frac{\phi_0}{2})\nonumber
\end{eqnarray}

In this region $|q^2/M_H^2|\ll 1$, and for the 
$ggH$-vertex $F^{\mu\nu}$ in the first order of the strong coupling we have the usual~\cite{SMHiggs} expression
\begin{eqnarray}
\label{ggHvertex}
F^{\mu\nu}&\!\!\simeq\!\!&\left( g^{\mu\nu}-\frac{2q_2^{\mu}q_1^{\nu}}{M_H^2}\right)
F_{gg\to H}\; ,\\
F_{gg\to H}&\!\!=\!\!& M_H^2\frac{\alpha_s}{2\pi}\sqrt{\frac{G_F}{\sqrt{2}}}f(\eta)\;,\\
f(\eta)&\!\!=\!\!&\frac{1}{\eta}\left\{ 1+\frac{1}{2}\left( 1-\frac{1}{\eta} \right) \left[ 
Li_2\left( \frac{2}{1-\sqrt{1-\frac{1}{\eta}}-i0} \right)+
Li_2\left( \frac{2}{1+\sqrt{1-\frac{1}{\eta}}+i0} \right)
\right]
\right\}\; ,\nonumber
\end{eqnarray}
where $\eta=M_H^2/4 m_t^2$, $G_F$ is the Fermi constant. NLO K-factor 1.5 for the $gg\to H$ process 
is included to the final answer.

 Taking the general form for $T$-amplitudes that satisfy identities
\begin{equation} 
\label{wardid}
q^{\alpha}T^D_{\mu\alpha}=0,\; q_i^{\mu}T^D_{\mu\alpha}=0\;, 
\end{equation}
and neglecting terms of the order
$o(\xi_i)$, the following expression is found at $|t_i|\le 1$~GeV$^2$:
\begin{equation}
\label{Tamp}
T^{D}_{\mu\alpha}(p,\; q,\; q_i)=\left( G_{\mu\alpha} - \frac{P^{q_i}_{\mu}P^{q}_{\alpha}}{P^{q_i}P^q} 
\right) T^{D}_{gp\to gp}(s_i,t_i,qq_i)
\end{equation}
$$
\noindent G_{\mu\alpha}=g_{\mu\nu}-
\frac{q_{i,\mu}q_{\alpha}}{qq_i}\; ,
$$
$$
\noindent P^{q_i}_{\mu}=p_{\mu}-\frac{pq_i}{qq_i}q_{\mu}\; ,
$$
$$
\noindent P^{q}_{\alpha}=p_{\alpha}-\frac{pq}{qq_i}q_{i,\alpha}\; .
$$
For $T^{D}_{gp\to gp}$ we use the
Regge-eikonal approach~\cite{i2,3Pomerons}. At small $t_i$ it takes the form
of the Born approximation, i.e. Regge factor: 
\begin{equation}
\label{REapproach}
T^D_{gp\to gp}(s_i,t_i,qq_i)= c_{gp}
\left(e^{-i\frac{\pi}{2}}
\frac{s_i-qq_i-m^2}{s_0-qq_i-m^2}
\right)^{\alpha_P(t_i)}
e^{b_0t_i}\;,
\end{equation}
$$
b_0=\frac{1}{4}(\frac{r^2_{pp}}{2}+r^2_{gp})\;,
$$
where $\alpha_P(0)=1.203$, $\alpha_P^{\prime}(0)=0.094$~GeV$^{-2}$, $r^2_{pp}=2.477$~GeV$^{-2}$ are found
in~\cite{3Pomerons}. Parameters $c_{gp}\simeq 3.5$, $r^2_{gp}=2.54$~GeV$^{-2}$ are defined from fitting 
the HERA data on elastic $J/\Psi$ production, which will be published elsewhere. The upper bound for the 
constant $c^{up}_{gp}\simeq 2.3(3.3)$ can be also estimated from
the exclusive double diffractive di-jet production at Tevatron, if we take CDF cuts and the
upper limit for the exclusive total di-jet cross-section~\cite{CDF}. The effective value $c_{gp}=2.3$ corresponds 
to the case, when the Sudakov suppression factor is absorbed into the constant, and $c_{gp}=3.3$ is obtained 
when taking into account this factor explicitely.

 The full amplitude for Higgs boson production looks as follows:
\begin{equation}
\label{Tpp_pHp}
T_{pp\to pHp}\simeq
\int\frac{d^4 q}{(2\pi)^4}\frac{8 F^{\mu\nu}(q_1,q_2)T^{D}_{\mu\alpha}(p_1,\; q,\; q_1)T^{D}_{\nu\alpha}(p_2,\; -q,\; q_2)}{(q^2+i 0) (q_1^2+i 0) (q_2^2+i 0)}\; .
\end{equation}
Factor $8$ arises from the colour index contraction. Let $l^2=-q^2\simeq{\mbox{\bf q}}^2$, $y_H=<y_H>=0$ and
contract all the tensor indices, then the integral~(\ref{Tpp_pHp})
takes the form

\begin{equation}
\label{Tpp_pHp2}
T_{pp\to pHp}\simeq
c_{gp}^2 e^{b(t_1+t_2)}\frac{\pi}{(2\pi)^2}
\left(-\frac{s}{M_H^2}\right)^{\alpha_P(0)}\cdot 8 F_{gg\to H}\cdot I\; ,
\end{equation}
\begin{equation}
b= \alpha^{\prime}_P(0) \ln\left(\frac{\sqrt{s}}{M_H}\right)+b_0\; ,
\end{equation}
\begin{equation}
\label{II}
I\simeq \int_{0}^{M_H^2}\frac{dl^2}{l^4} 
\left(\frac{l^2}{s_0-m^2+l^2/2}\right)^{2\alpha_P(0)}\simeq 1.88\; \mbox{\rm GeV}^{-2}\;,
\end{equation}
where $M_H=100$~GeV, and $s_0-m^2\simeq 1$~GeV$^2$ is the scale parameter of the model that is used in the global fitting of 
the data on $pp(p\bar{p})$ scattering for on-shell amplitudes~\cite{3Pomerons}. It remains fixed in the present 
calculations. 

If we take into account the emission of virtual "soft" gluons, while prohibiting the real ones, that 
could fill rapidity gaps, it results in
the Sudakov-like suppression~\cite{Khoze}:

\begin{eqnarray}
\label{sudakov}
F_s(l^2)=exp\left[-\frac{3}{2\pi}\int\limits_{l^2}^{{M_H}^2/4}\frac{d{p_T}^2}{{p_T}^2}\alpha_s({p_T}^2)\ln\left(
\frac{{M_H}^2}{4{p_T}^2}\right)\right]\; ,
\end{eqnarray}
and to the new value of the integral~(\ref{II}):
\begin{equation}
\label{IInew}
I_s\simeq \int_{0}^{M_H^2}\frac{dl^2}{l^4} F_s(l^2) 
\left(\frac{l^2}{s_0-m^2+l^2/2}\right)^{2\alpha_P(0)}\simeq 0.38\; \mbox{\rm GeV}^{-2}\;, M_H=100\; GeV.
\end{equation}
In this case the total cross-section becomes $24$ times smaller, than without the factor $F_s$.

Unitarity corrections can be estimated from the elastic $pp$ 
scattering by the method depicted in Fig.1, where

\begin{eqnarray}
\label{ucorr}
T_X&=&T_{pp\to pHp}\;,\\
V(s\;,\;\mbox{\bf q}_T)&=&4s(2\pi)^2\delta^2(\mbox{\bf q}_T)
+4s\int d^2\mbox{\bf b}e^{i\mbox{\bf q}_T \mbox{\bf b}}
\left[e^{i\delta_{pp\to pp}}-1\right]\;,\nonumber\\
T^{Unit.}_X(p_1\;,\; p_2\;,\;\Delta_1\;,\;\Delta_2)
&=&\frac{1}{16ss^{\prime}}\int
\frac{d^2\mbox{\bf q}_T}{(2\pi)^2}\frac{d^2\mbox{\bf q}^{\prime}_T}{(2\pi)^2}
V(s\;,\;\mbox{\bf q}_T)\;
\cdot\;T_X( p_1-q_T, p_2+q_T,\Delta_{1T},
\Delta_{2T})\;\cdot\nonumber\\
&\cdot&\; V(s^{\prime}\;,\;\mbox{\bf q}^{\prime}_T)\;,\nonumber\\
\Delta_{1T}&=&\Delta_{1}-q_T-q^{\prime}_T\;,\nonumber\\
\Delta_{2T}&=&\Delta_{2}+q_T+q^{\prime}_T\;,\nonumber
\end{eqnarray}
and $\delta_{pp\to pp}$ can be found in Ref.~\cite{3Pomerons}. It reduces the integrated cross-section 
by the factor about $14$.

\section{Results and discussions} 

 We have the following expression for the differential cross-section:
\begin{eqnarray}
\label{dsigttxx}
\frac{d\sigma}{dt_1dt_2d\xi_1d\xi_2}\!\!\!\!\!\!\!\!\!\!\!\!& &=\frac{\pi |T^{Unit.}_{pp\to pHp}|^2}{8s(2\pi)^5\sqrt{-\lambda}}\\
&\lambda&=\kappa^2+2(t_1+t_2)\kappa+(t_1-t_2)^2\le 0\nonumber
\end{eqnarray}
 
 By partial integrating~(\ref{dsigttxx}) we obtain the cross-sections $d\sigma/dt$ and $d\sigma/d\xi$. The first 
result of our calculations is depicted 
in the Fig.~3. The antishrinkage of the 
diffraction peak with
increasing Higgs boson mass is the direct consequence of the existence of the additional
hard scale $M_H$, which makes the interaction radius smaller. The $\xi$ dependence
is shown in Fig.4.

 The second result is the total cross-section versus Higgs boson mass. We obtain the 
following results for the total cross-section of the EDD Higgs boson production:

\vspace*{0.4cm}
\begin{tabular}{|c|c|c|c|c|c|}
\hline
          &                 &\multicolumn{4}{|c|}{$\sigma_{p+p\to p+H+p}$ (fb)}\\
\cline{3-6} 		  
$c_{gp}$  &  $M_H$ (GeV)    &\multicolumn{2}{|c|}{LHC}      &\multicolumn{2}{|c|}{TeVatron}\\
\cline{3-6}
          &                 & no Sud. suppr.& Sud. suppr.     & no Sud. suppr.   & Sud. suppr.\\
\hline 
   3.5  &  $100\to 500$   &  $110\to 57$  & $4.6\to 0.14$  & $12\to 0.4$   & $0.5\to 0.001$ \\
\hline 
   2.3(3.3) &  $100\to 500$   &  $20\to 11$ & $\mbox{\bf 3.6}\to \mbox{\bf 0.11}$  & $2.2\to 0.08$     & $0.4\to 0.0009$  \\
\hline 
\end{tabular}
\vspace*{0.4cm}

 The above results for LHC energies are depicted in Figs.5-8. The form of the curve originates from the
standard gluon-gluon-Higgs vertex and has a peak near $M_H\simeq 2m_t$. For the case of Sudakov-like
suppression the cross-section vanishes faster with $M_H$.

It is useful to compare our result with other studies. Results quite close to ours (with the normalization to
the CDF data, $c_{gp}=3.3$) 
were given in Ref.~\cite{Khoze}, where the value of the total cross-section is 
about $3$~fb. In both cases the most important suppressing in the mass region $M_H>100$~GeV is due to
(perturbative) Sudakov factors, while the nonperturbative (absorbtive) factors play relatively
minor role. 

 Results of other authors were considered in details in~\cite{Khozemyths}. The highest cross-section $2$~pb for $M_H=400$~GeV 
at LHC energies was obtained in Ref.~\cite{Cudell}. They used a nonfactorized
form of the amplitude and a "QCD inspired" model for $g p\to g p$ amplitudes, taking into account
the nonperturbative proton wave functions. Even if we multiply the result of Ref.~\cite{Cudell} by 
the suppressing factor, it will be larger than ours. This could serve as the indication of
the role of nonperturbative effects. Our model is based on the Regge-Eikonal approach for
the amplitudes, which is primordially nonperturbative, normalized to the 
data from HERA on $\gamma p\to J/\Psi p$ and improved by the CDF data on the exclusive di-jet production.  

 To estimate the signal to QCD background ratio 
for $b\bar{b}$ signal we use the standart expression for $gg\to b\bar{b}$ amplitude and assumptions~\cite{Khozebg1chi}-\cite{KhozeJz0}:
\begin{itemize}
\item possibility to separate final $b\bar{b}$ quark jets from gluon jets. If we cannot
do it, it will increase the background by two orders of magnitude under the $50\%$ efficiency.
\item suppression due to the absence of colour-octet $b\bar{b}$ final 
states 
\item suppression of light fermion pare production, when $J_{z,tot}=0$ (see also~\cite{Jz0lvanish1},\cite{Jz0lvanish2})
\item cut $E_T>50$~GeV ($\theta\ge 60^{\mbox{\small o}}$), since the cross-section of
diffractive $b\bar{b}$ jet production strongly decreases with $E_T$.    
\end{itemize}
The theoretical result of these numerical estimations is
$$
\frac{Signal(pp\to pHp\to pb\bar{b}p)}{QCD\;\;background}\ge
$$
\begin{equation}
\label{sigbgen}
\ge 9.5\cdot 10^{-6} |f|^2 Br_{H\to Q\bar{Q}}\frac{M_H^3}{\Delta M}\;{,}
\end{equation}
where $|f|^2\simeq 0.5\to 3$ for Higgs boson masses $100\to 350\; GeV$. For 
$M_H\simeq 115$~GeV 
\begin{equation}
\label{sigbgex}
\frac{Signal(pp\to pHp\to pb\bar{b}p)}{QCD\;\;background}\ge 3.8 \frac{GeV}{\Delta M}\;{,} 
\end{equation}
where $\Delta M$ is the mass resolution of the detector. Similar result was strictly obtained 
in~\cite{Khozebg1chi},\cite{Khozebg2full}. More exact estimations of the above ratio by fast Monte-Carlo simulations and the 
total efficiency of EDD Higgs boson production will be publised in a forthcoming paper.


\section{Conclusions}

 We see from the result that there is a real possibility to detect Higgs boson using
the usual $b\bar{b}$ signal under the luminosity greater than $10^{32}$ in EDD events
at LHC. Accuracy of the mass measurements could be improved by applying the missing mass method~\cite{Rostovtsev}.

 The low value of the exclusive Higgs boson production cross-section obtained in this paper is mainly due to
the Sudakov suppression factor~(\ref{sudakov}), the full validity of which is not obvious for us, because
the confinement effects can strongly modify the "real gluon emission".

 It is interesting that in spite of different models and quite different ways of account of 
absorbtive effects in our paper and in Ref.~\cite{Khoze}, the final results (see Fig.8) appeared to be quite close.

 Certainly, the cross-sections may be several times lager due to still not 
very well known non-perturbative factors.

 It is possible to generalize our approach for the exclusive production of
other particles like $\chi_{c0}$, $\chi_{b0}$, radion, K.-K. gravitons, glueballs. In this case
cross-sections can be lager than for EDD Higgs boson production, and some other important investigations like
partial wave analysis and measurements of the diffractive pattern of the interaction could be done.
 
\section*{Aknowledgements}

 We are grateful to A. De Roeck, A. Prokudin, A. Rostovtsev, A. Sobol, S. Slabospitsky and participants of
BLOIS2003 workshop and several CMS meetings for helpful discussions. This work is supported by the Russian
Foundation for Basic Research, grant no. 02-02-16355




\newpage
\section*{Figure captions}

\begin{list}{Fig.}{}

\item 1: The full unitarization of the process $p+p\to p+X+p$.
\item 2: The process $p+p\to p+H+p$. Absorbtion in the initial and final pp-channels is not shown.
\item 3: t-distribution $d\sigma/dt/\sigma_{tot}$ of the process $p+p\to p+H+p$ for Higgs boson masses 100 and 500 GeV.
\item 4: $\xi$-distribution $d\sigma/d\xi/\sigma_{tot}$ of the process $p+p\to p+H+p$ for $M_H=100$~GeV.
\item 5: The total cross-section (in fb) of the process $p+p\to p+H+p$ versus Higgs boson mass for $c_{gp}=3.5$ 
without Sudakov-like suppression.
\item 6: The total cross-section (in fb) of the process $p+p\to p+H+p$ versus Higgs boson mass for $c_{gp}=3.5$ 
with Sudakov-like suppression.
\item 7: The total cross-section (in fb) of the process $p+p\to p+H+p$ versus Higgs boson mass for $c_{gp}=2.3$ 
without Sudakov-like suppression.
\item 8: The total cross-section (in fb) of the process $p+p\to p+H+p$ versus Higgs boson mass for $c_{gp}=3.3$ 
with Sudakov-like suppression.

\end{list}


\newpage

\begin{figure}[hb]
\label{unitar}
\vskip 4cm
\hskip  1cm \vbox to 14cm {\hbox to 16cm{\epsfxsize=16cm\epsfysize=14cm\epsffile{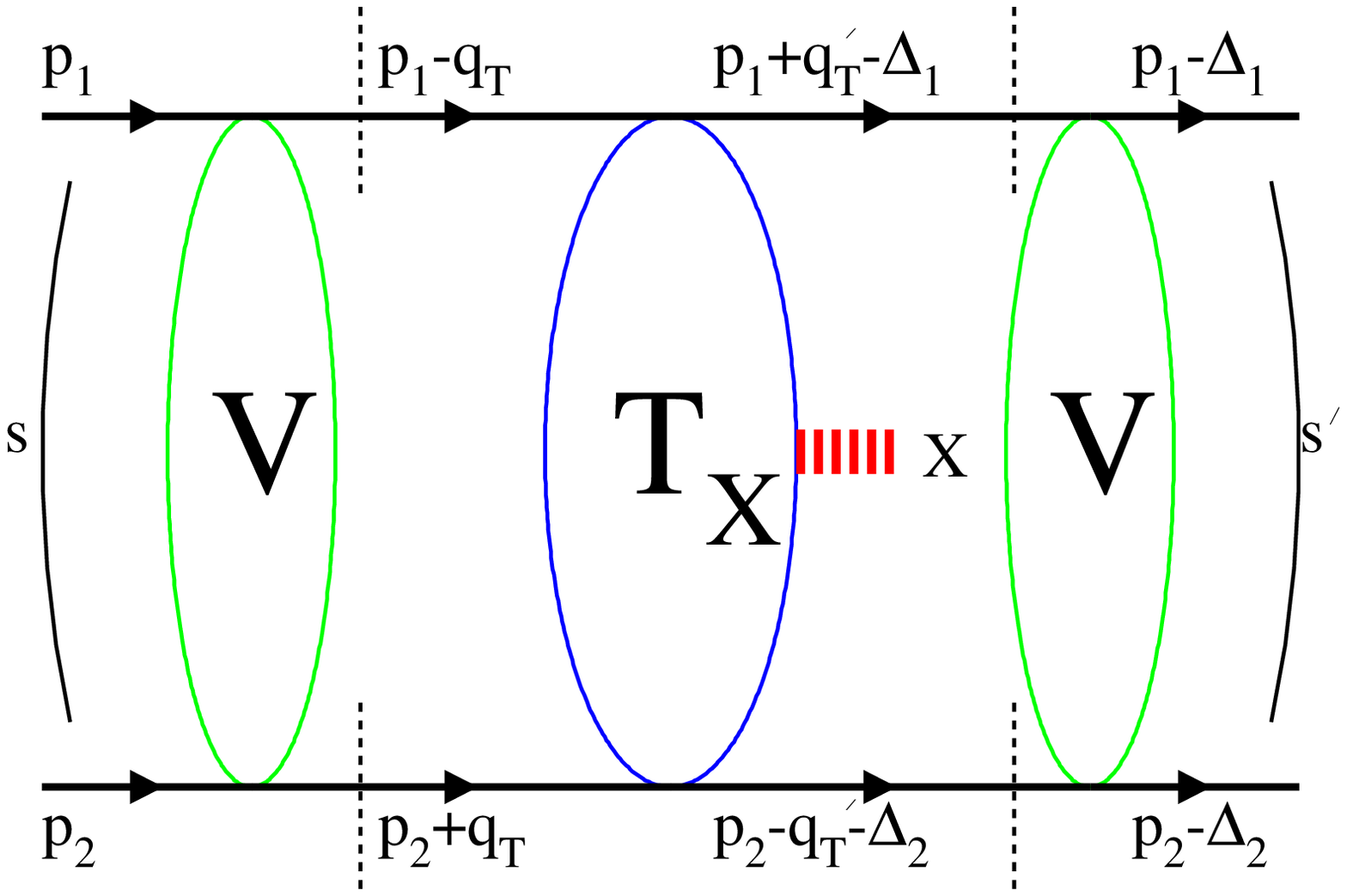}}}
\hskip 1cm
\caption{}
\end{figure}

\newpage

\begin{figure}[hb]
\label{pp_pHp}
\vskip 1.5cm
\vbox to 15cm {\hbox to 15cm{\epsfxsize=15cm\epsfysize=15cm\epsffile{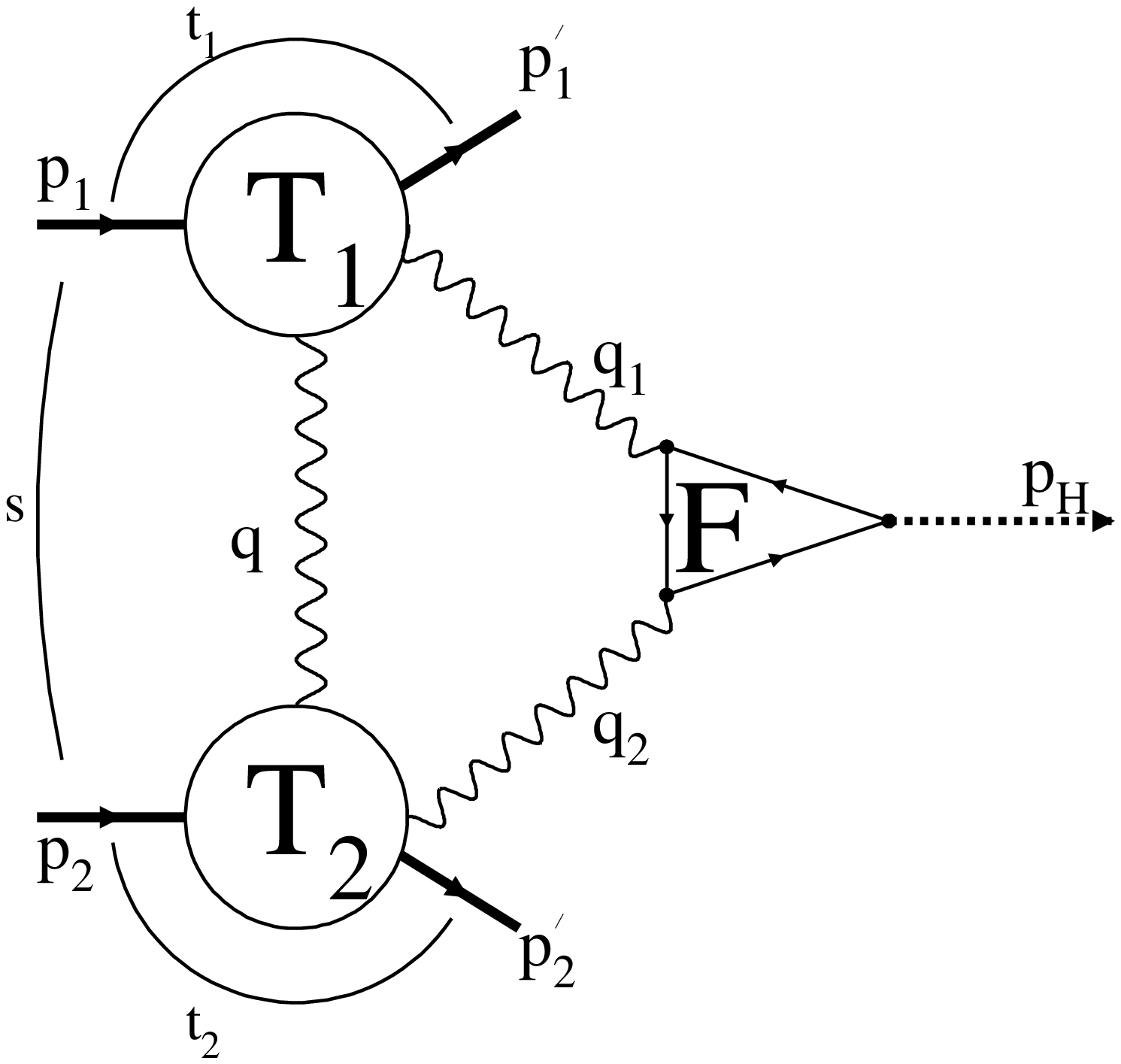}}}
\hskip 1cm
\caption{}
\end{figure}

\newpage

\begin{figure}[hb]
\label{dsigtHiggs}
\vskip 1.5cm
\vbox to 17cm {\hbox to 17cm{\epsfxsize=17cm\epsfysize=17cm\epsffile{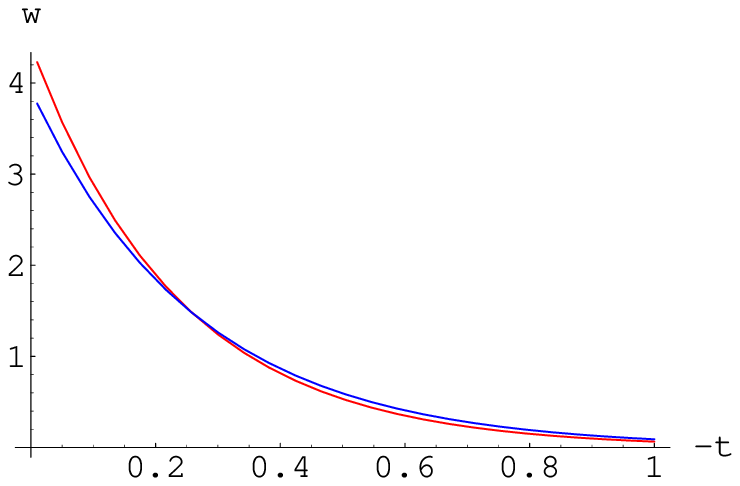}}}
\hskip 1cm
\caption{}
\end{figure}

\newpage

\begin{figure}[hb]
\label{dsigxHiggs}
\vskip 1.5cm
\vbox to 17cm {\hbox to 17cm{\epsfxsize=17cm\epsfysize=17cm\epsffile{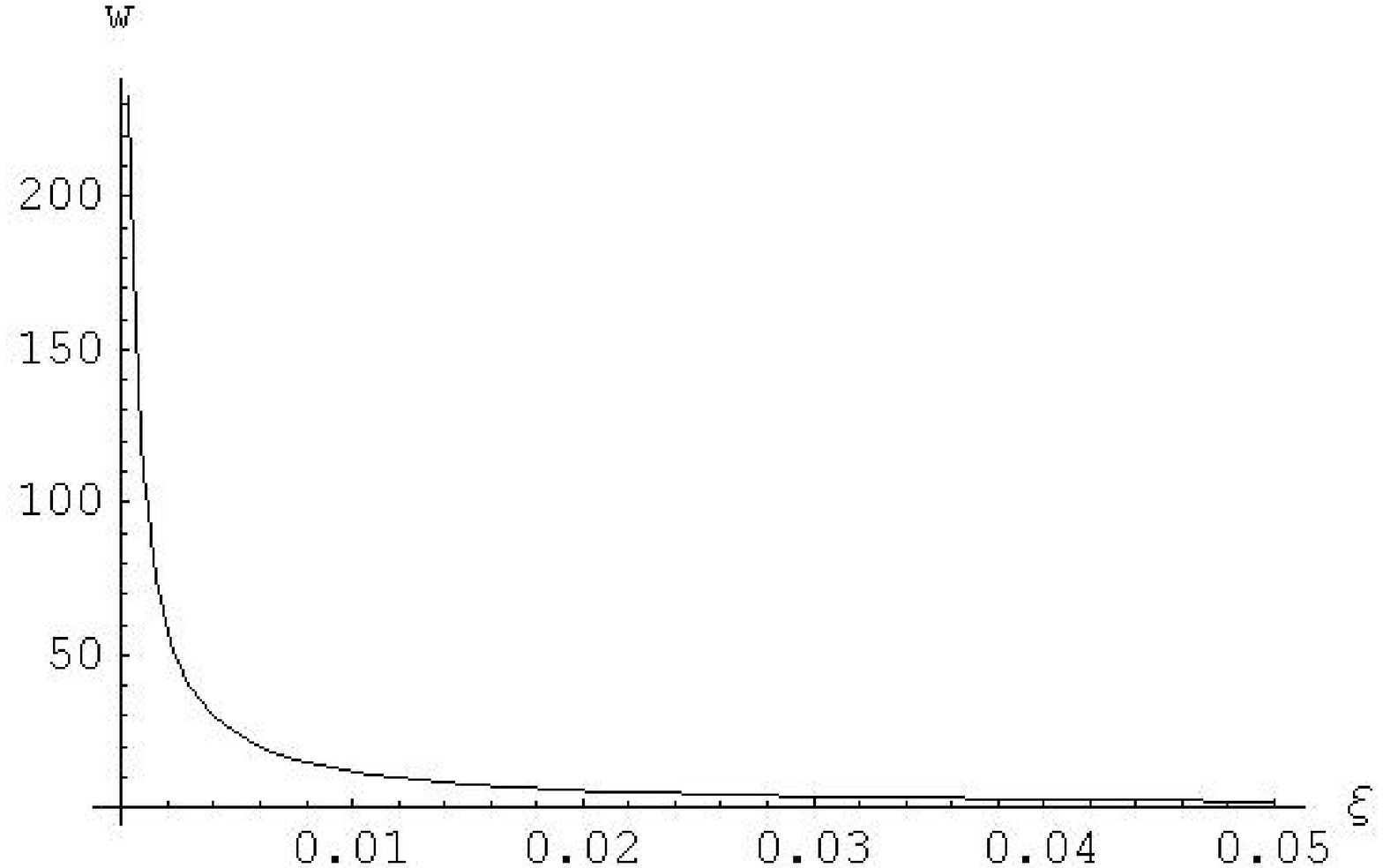}}}
\hskip 1cm
\caption{}
\end{figure}

\newpage

\begin{figure}[hb]
\label{sigtotLHC0}
\vskip 1.5cm
\vbox to 17cm {\hbox to 17cm{\epsfxsize=17cm\epsfysize=17cm\epsffile{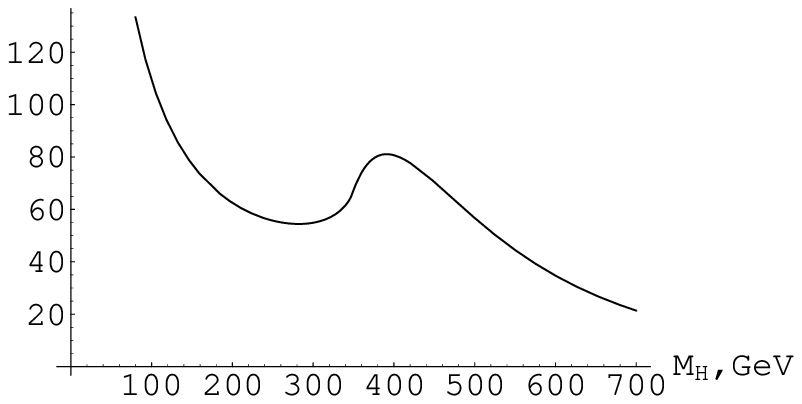}}}
\hskip 1cm
\caption{}
\end{figure}

\newpage

\begin{figure}[hb]
\label{sigtotLHC0su}
\vskip 1.5cm
\vbox to 17cm {\hbox to 17cm{\epsfxsize=17cm\epsfysize=17cm\epsffile{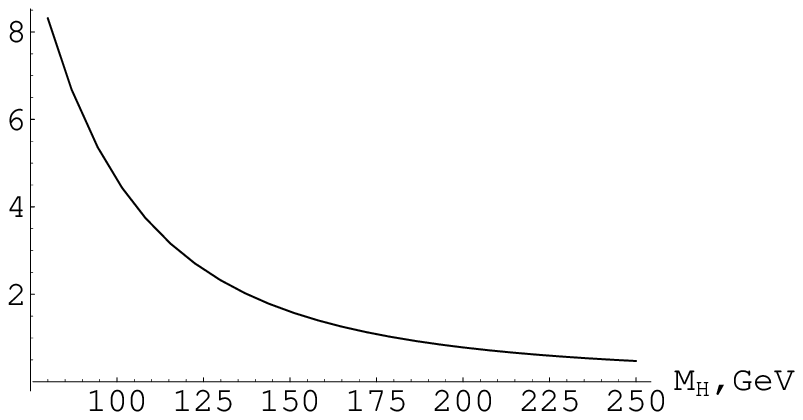}}}
\hskip 1cm
\caption{}
\end{figure}

\newpage

\begin{figure}[hb]
\label{sigtotLHCup}
\vskip 1.5cm
\vbox to 17cm {\hbox to 17cm{\epsfxsize=17cm\epsfysize=17cm\epsffile{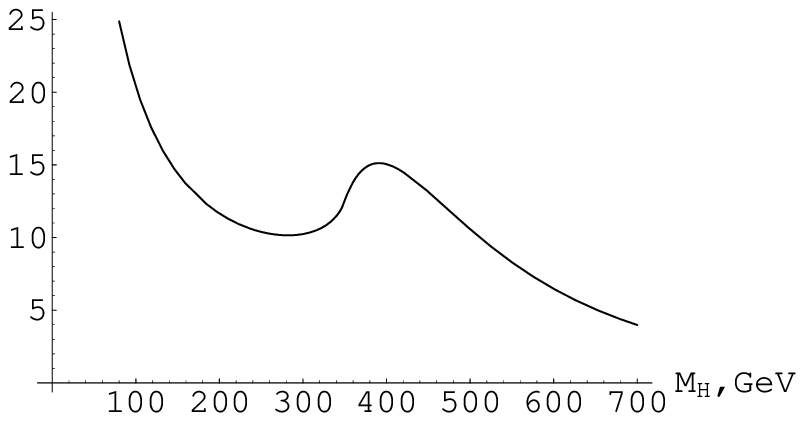}}}
\hskip 1cm
\caption{}
\end{figure}

\newpage

\begin{figure}[hb]
\label{sigtotLHCupsu}
\vskip 1.5cm
\vbox to 17cm {\hbox to 17cm{\epsfxsize=17cm\epsfysize=17cm\epsffile{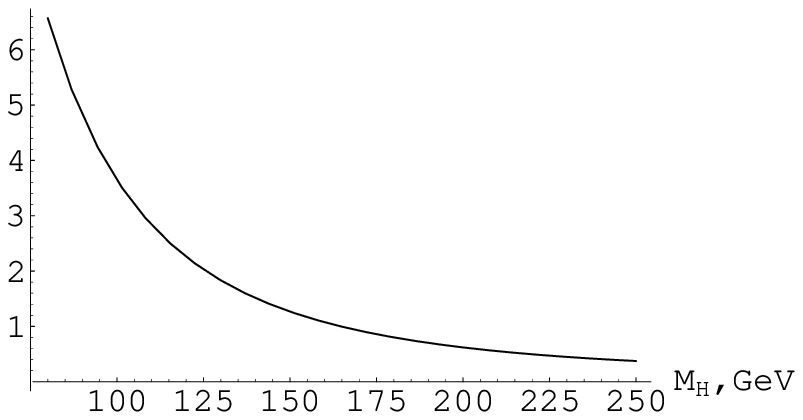}}}
\hskip 1cm
\caption{}
\end{figure}

\end{document}